\documentclass[preprint2]{aastex}
\input psfig.tex

\begin{document}

\title {Color-magnitude relations of late-type galaxies}

\author{Ruixiang Chang \altaffilmark{1}; Shiyin Shen \altaffilmark{1};
 Jinliang Hou \altaffilmark{1}; Chenggang Shu \altaffilmark{2,1};
 Zhengyi Shao \altaffilmark{1}}

\altaffiltext{1}{Shanghai Astronomical Observatory, 80 Nandan
Road, Shanghai, China, 200030} \altaffiltext{2}{Shanghai Normal
University,  Road, Shanghai, China, 200030}

\begin{abstract}
We use a large sample of galaxies drawn from the Sloan Digital Sky
Survey (SDSS) and Two Micro All Sky Survey (2MASS) to present
Color-Magnitude Relations (CMRs) for late-type galaxies in both
optical and optical-infrared bands. A sample from SDSS Data
Release 4 (DR4) is selected to investigate the optical properties.
Optical-infrared colors are estimated from a position matched
sample of DR4 and 2MASS, in which the photometric aperture
mismatch between these two surveys is carefully corrected. It is
shown that, after correcting the dust attenuation, the optical
colors for faint galaxies (i.e. $M_r > -21$) have very weak
correlation with the luminosity, while the optical colors for
bright galaxies (i.e. $M_r < -21$) are redder for more luminous
galaxies. All (optical, optical-infrared and infrared) colors show
similar but stronger correlations with stellar mass than with
absolute magnitude. The optical colors correlate more strongly
with stellar mass surface density than with stellar mass, while
optical-infrared and infrared colors show stronger correlations
with stellar mass. By comparing the observed colors of our sample
galaxies with the colors predicted by stellar population synthesis
model, we find that massive late-type galaxies have older and
higher metallicity stellar population than less massive galaxies.
This suggests that CMRs for late-type galaxies are trends defined
by the combination of stellar mean age and metallicity. Moreover,
our results suggest that the stellar mean metallicity of late-type
galaxy is mainly determined by its stellar mass, while the star
formation history is mainly regulated by the stellar mass surface
density.
\end{abstract}

\keywords{Galaxies:evolution -- Galaxies:photometry -- Galaxies:
stellar content}

\section{Introduction}

It has long been known that early-type galaxies in nearby clusters
exhibit a tight Color-Magnitude Relation (CMR), i.e. more luminous
early-type galaxies have redder colors than less luminous ones
(Baum 1959; Visvanathan \& Sandage 1977; Bower et al. 1992), and
this finding has been confirmed by later studies (Aragon-Salamanca
et al. 1993; Stanford et al. 1995, 1998; Ellis et al. 1997; Kodama
et al. 1998; Terlevich et al. 2001; Bernardi et al. 2003, 2005;
Blanton et al. 2003a; Bell et al. 2004; Holden et al. 2004; Hogg
et al. 2004; Cool et al. 2006; Chang et al. 2006). This relation
has been mainly ascribed to the metallicity effect, either because
more massive galaxies have stronger binding energies so that the
gas can be enriched to higher metallicity (Faber 1977; Dressler
1984; Arimoto \& Yoshii 1987; Kodama \& Arimoto 1997), or because
larger ellipticals form at low redshift through major mergers
between more massive, metal-rich disk systems (Kauffmann \&
Charlot 1998; De Lucia et al. 2004; Kang et al. 2005).

However, the case for late-type galaxies is much more complicated
partially because of the influence of on-going star formation and
the effect of dust attenuation. Almost thirty years ago,
Visvanathan \& Griersmith (1977) find that early-type spiral
galaxies (S0/a to Sab) in the Virgo cluster exhibit very similar
optical CMR to that of E/S0 galaxies, i.e. brighter spirals show
redder colors, except that the former has a much larger scatter.
Based on a large sample from Sloan Digital Sky Survey (SDSS, York
et al. 2000), Blanton et al. (2003b) investigate the broadband
optical properties for concentrated galaxies (fitted Sersic index
$n>3$) and exponential galaxies ($n<1.5$) and find that
exponential galaxies have a less tight, but undeniable optical
CMR. Baldry et al. (2004) analyze the bimodal color-magnitude
distribution of SDSS galaxies and find that both red and
blue-distribution galaxies can be well fitted by a straight line
plus a \emph{tanh} function. In other words, the low-luminosity
blue-distribution galaxies ($M_r>-19$) show a shallower CMR slope
than high-luminosity galaxies. However, these works are all
limited to the optical bands.

Tully et al. (1998) investigate global extinction in spiral
galaxies and establish optical-IR CMRs for spirals after
correction for dust attenuation. Peltier \& de Grijs (1998)
determine an optical-infrared CMR ($I-K$ vs. $M_K$) for spiral
galaxies. They find the CMR is very tight and its slope is steeper
than that for elliptical and S0 galaxies. This result is
consistent with early studies  (e.g. Wyse 1982; Tully et al. 1982;
Mobasher et al. 1986). Using stellar population synthesis (SPS)
models, they conclude that faint spiral galaxies have both a
younger age and a lower metallicity than bright spirals. Bell \&
de Jong (2000) have compiled a sample of 121 low-inclination
spiral galaxies with radially resolved optical and near-infrared
photometry. By comparing the available photometry with the SPS
model colors, they found that the star formation history of a
galaxy (as probed by its age, metallicity and gas fraction)
strongly correlates with its surface brightness and magnitude.
These results are confirmed by Macarthur et al. (2004). However,
all those early studies are limited to small galaxy samples since
it is difficult to get large and homogeneous optical and infrared
photometric samples simultaneously.

On the other hand, the interpretation of these observed CMRs is
still a challenge for semi-analytic models. van den Bosch (2002)
and Bell et al. (2003) find that there is significant discrepancy
between the predictions of semi-analytic models and the data. The
observations show that brighter spiral galaxies are redder than
fainter ones, while the models predict that fainter spirals should
be slightly redder than bright spirals. Recently, Kang et al.
(2005) introduce a semi-analytic model based on high-resolution
N-body simulations and compare their model predictions with the
observed bimodal distribution in the color-magnitude diagram found
by Baldry et al. (2004). They find that the prediction for the red
branch is very close to the observations, but for the blue part,
the predicted colors of luminous spirals are bluer and those of
less luminous spirals are redder than observations. All these
results suggest that the implied physics of gas accretion and
feedback during the formation and evolution of spiral galaxies are
still not well understood (van den Bosch 2000; Bell et al. 2003).

It should be pointed out that the comparison presented by Kang et
al.(2005) is only limited to the color $(u-r)$, which is very
sensitive to dust attenuation. Furthermore, the degeneracy between
stellar age and metallicity for optical colors complicates further
investigations for the star formation history of late-type
galaxies. Thus, there is still a long way to go before getting
robust conclusions. Since the combination of optical and
near-infrared broad-band colors can partially break
age-metallicity degeneracy, it is important to establish
observationally the CMRs in both optical and near-infrared bands
for late-type galaxies based on a large homogenous sample.

The SDSS and Two Micron All Sky Survey (2MASS, Jarrett 2000a,b)
have already created the largest well-defined sample of galaxies
with well-measured optical and infrared photometry to date.
Recently, we have analyzed the optical and infrared color
properties for early-type galaxies after correcting the aperture
mismatch between these two magnitude systems (Chang et al. 2006).
This paper is a complementary study of Chang et al. (2006). Here
the samples are selected for late-type galaxies. One of our aims
is to establish optical, optical-infrared and infrared CMRs for
late-type galaxies by using the largest homogeneous samples
up-to-date.

To derive the optical-infrared colors, we adopt the method
presented in Chang et al. (2006) and correct SDSS magnitudes to
the aperture where 2MASS magnitudes are measured. For the purpose
of minimizing the different dust attenuation caused by different
inclinations, the galaxies in our sample are limited to face-on
spirals. We also calculate the emission-line-free colors by
comparing the magnitude difference before and after removing the
emission lines from the spectra. With these carefully corrected
optical and optical-infrared colors, we compare the correlations
between colors and absolute magnitude, stellar mass and stellar
mass surface density and try to distinguish which parameter mainly
determines the color properties for late-type galaxies. Moreover,
by comparing the colors of our sample galaxies with those of SPS
models, we investigate the stellar populations of spiral galaxies
and their implications on star formation processes in late-type
galaxies.

The outline of this paper is as follows. Section 2 describes the
way of selecting sample galaxies and the photometric quantities
used in the paper. Section 3.1 presents the method to do the
photometric aperture corrections. The magnitude corrections for
emission-line contamination is shown in Section 3.2. Section 4 is
the main part of the paper, where we present various correlations
between colors and absolute magnitudes, stellar mass and stellar
mass surface density and their comparisons. Section 5 describes
the SPS models and the model predictions for stellar populations
of spiral galaxies. Finally, in Section 6 we summarize the main
results of this paper.

\section{The data and samples}

The data used in this paper is mainly taken from Data Release 4 of
the SDSS (York et al. 2000; Stoughton et al. 2002, hereafter EDR;
Adelman-McCarthy et al. 2006). In particular, we use the Main
Galaxy Sample in the New York University Value-Added Galaxy
Catalog (NYU-VAGC), which is based on the SDSS database, but with
an independent set of enhanced reductions (Blanton et al. 2005).
From this catalog, we take spectroscopic redshift and photometric
quantities, including Petrosian magnitudes, K-corrections, axis
ratio $(b/a)$ from the exponential fitting ($a$ and $b$ are the
semimajor and semiminor axes) and $fracDev_r$. The parameter
$fracDev_r$ is obtained by taking the best-fit exponential and de
Vaucouleurs fit to the surface brightness profile, finding the
linear combination of the two that best fits the image, and
storing the fraction contributed by the de Vaucouleurs fit. We
also take the quantities from ProfMean, the azimuthally averaged
surface brightness in a series of fixed circular annuli (see table
7 in EDR), which are used to make the aperture correction for the
optical-infrared colors (see section 3.1).

The 2MASS is a ground-based, near-infrared imaging survey of the
whole sky and its Extended Source Catalog (XSC) contains nearly
1.6 million galaxies (Jarrett et al. 2000a,b). The infrared
magnitudes are taken from extended source catalog (XSC) of 2MASS.
In section 3.1, we will describe how to estimate the
optical-infrared colors.

Conversion from apparent magnitude $m$ to absolute magnitude $M$
depends on the adopted cosmology and the K-correction:
\begin{equation}
M=m-5 \log [ D_L(z,\Omega_M,\Omega_{\Lambda})]-25-K(z),
\end{equation}
where $D_L$ is the luminosity distance in unit of Mpc and $K(z)$
is the K-correction. We assume the standard cosmology with
$\Omega_M=0.3$, $\Omega_{\Lambda}=0.7$ and $H_0$=70 km s$ ^{-1}$
Mpc$^{-1}$ to calculate luminosity distance. We use SDSS $r-$band
Petrosian magnitude and calculate $M_r$ at redshift $z=0$.
K-corrections for magnitudes in the \it g, r, i, z \rm bands (and
\it J, H, K \rm bands if needed) are taken from NYU-VAGC (Blanton
et al. 2003a; Blanton et al. 2005). Stellar masses of galaxies are
taken from http://www.mpa-garching.mpg.de/SDSS (see details in
Kauffmann et al. 2003a).

\begin{figure} [t]
\psfig{file=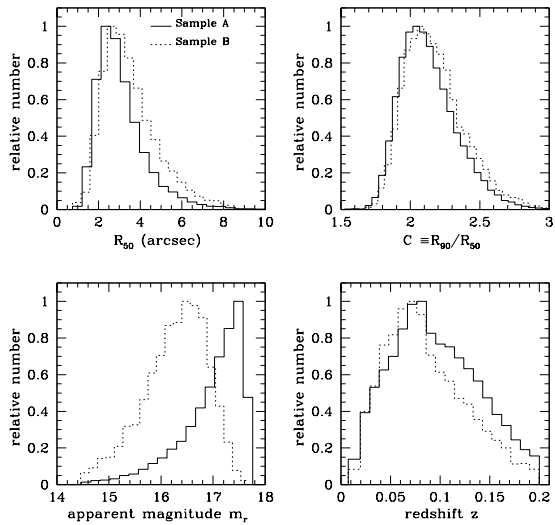,height=8.cm,width=0.5\textwidth}
 \caption[]{Distributions of several quantities for the galaxies
            in Sample A (solid lines) and Sample B (dotted lines),
            including $R_{50} (arcsec)$, concentration $C =
            R_{90}/R_{50}$, r-band apparent magnitude $m_r$ and
redshift $z$. }
 \label{ }
\end{figure}

\begin{figure} [t]
\psfig{file=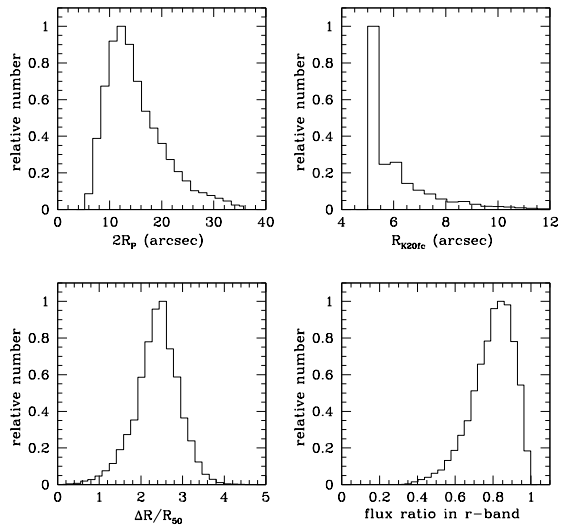,height=8.cm,width=0.5\textwidth}
\caption{Histograms of several quantities relevant to do aperture
corrections for galaxies in Sample B, including two times
Petrosian radius in the $r$-band $2R_P$ (from SDSS), the isophotal
radius $R_{k20fc}$ (from 2MASS), $\Delta R/R_{50}$ and the
$r$-band flux within $R_{k20fc}$ divided by the Petrosian flux.}
\end{figure}

Sample galaxies are selected from Main Galaxy Sample in the
NYU-VAGC (Blanton et al. 2005). We adopt the $r-$band morphology
indicator $fracDev_r$ to select late-type galaxies with criteria
$fracDev_r<0.5$, which means that the surface brightness of
selected galaxy is more likely to be expressed by exponential
profile. Adopting $fracDev$ parameter has the advantage of seeing
and inclination correction during the profile fitting (Bernardi et
al. 2005). From this criteria, our sample contains 154,870
late-type galaxies.

We further restrict our sample with the condition $b/a>0.75$ so as
to select nearly face-on galaxies based on two considerations.
Firstly, edge-on galaxies are more dust-attenuated and reddened
than face-on systems (Tully et al. 1998; Xiao et al. 2006). This
effect will introduce extra dispersion into the CMRs. In other
words, by selecting face-on galaxies, we can minimize the effects
of dust attenuation caused by different inclination of the spiral
disks. Secondly, the isophoto aperture of the high axis-ratio
galaxies is nearly circular. In this case, the SDSS ProfMean
measured in circular radii can be properly used to make the
aperture correction of optical-infrared colors with enough
accuracy. After this, the number of late-type galaxies is reduced
to 45,033.

Furthermore, we restrict our sample to galaxies in the redshift
range $0.01<z<0.2$. The lower limit is to reduce the uncertainty
using the redshift as distance estimator. The upper limit is to
lower the uncertainties on K-corrections for the magnitudes and to
reduce the possible color evolution with redshift. Finally, the
number of galaxies is further reduced to 40,987. We denote this
sample as Sample A and use it to investigate the optical CMRs.

The NYU-VAGC also contains a matched sample between SDSS and 2MASS
XSC. This sample is obtained by positionally matching sources
observed by 2MASS to galaxies in the `main' spectroscopic sample
of the SDSS within $3 arcsec$. We select all the galaxies in
sample A with 2MASS match to create Sample B. There is no cut in
neither $m_K$ nor surface brightness in K-band. The final number
of galaxies in Sample B is 9,930. Sample B is used to investigate
the optical-infrared and infrared CMRs.

Fig. 1 plots histograms of several quantities for the galaxies in
Sample A (solid lines) and Sample B (dotted lines), including
$R_{50} (arcsec)$, concentration $C \equiv R_{90}/R_{50}$
($R_{90}$ and $R_{50}$ are the radii including 50\% and 90\% of
the Petrosian fluxes), r-band apparent magnitude $m_r$ and
redshift $z$. As we can see from the top-right panel of Fig. 1,
most of the sample galaxies have $C<2.6$, consistent with the
early studies of the morphology separation with parameter $C$
(e.g. Strateva et al. 2001). Compared with Sample A, the galaxies
in Sample B are biased to be apparently bright, large and to lower
redshifts. This is due to the fact that 2MASS is much more shallow
than SDSS.

\section{Corrections for magnitudes}

\subsection{Aperture corrections}\label{Apert_corr}

In order to estimate the optical-infrared colors, it is important
that both SDSS and 2MASS magnitudes are measured within the same
aperture. We adopt the same method as that in Chang et al. (2006)
to do aperture corrections for the magnitudes, i.e. to correct
SDSS magnitudes to the aperture where the 2MASS magnitudes are
measured.

\begin{figure}[t]
\psfig{file=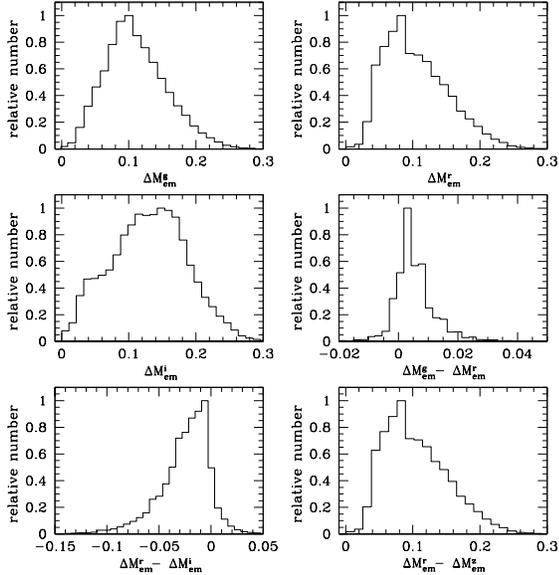,height=8.cm,width=0.5\textwidth}
\caption{Histograms of emission-line corrections for magnitudes in
$g-$band (left-top), $r-$band (right-top), $i-$band (left-meddle)
and colors $g-r$ (right-middle), $r-i$ (left-bottom) and $r-z$
(right-bottom) in the fiber.}
\end{figure}

There are several sets of magnitudes provided in the 2MASS XSC. In
this paper, we use the isophotal fiducial magnitudes, which are
measured within the circular aperture corresponding to a surface
brightness of 20.0 mag arcsec$^{-2}$ in the $K_s$-band (the
aperture is denoted as $R_{K20fc}$). SDSS provides the azimuthally
averaged surface brightness in a series of circular annuli, i.e.
ProfMean. We calculate $g, r, i, z$ magnitudes that are matched to
the 2MASS measurements by interpolating the cumulative radial
surface brightness profile in each SDSS band at the corresponding
isophotal radius $R_{K20fc}$ (using the recommended approach in
section 4.4.6.2 of EDR). The derived magnitude in \it r-\rm band
$m_{K}^r$ is used to estimate the optical-infrared colors
throughout this paper. The average value of magnitude correction
in the r-band for aperture mismatch is 0.26.

Fig. 2 plots the distribution of several quantities for galaxies
in Sample B, including two times Petrosian radius in the $r$-band
$2R_P$ (from SDSS)\footnote{The $r$-band $2R_P$ is the aperture to
which the Petrosian magnitudes are integrated, see EDR}, the
isophotal radius $R_{K20fc}$ (from 2MASS), the aperture mismatch
$\Delta R /R_{50,r}= (2R_P-R_{K20fc})/R_{50,r}$ and the ratio
between the $r$-band flux within $R_{K20fc}$ and the Petrosian
flux. It can be seen that the aperture mismatch between SDSS and
2MASS is quite large. However, more than half of the Petrosian
flux is already contained within $R_{K20fc}$ for most of galaxies.

\subsection{Emission-line corrections}

Since late-type galaxies have on-going star formation, emission
lines from HII regions have significant contributions to the
observed magnitudes. However, SPS models generally do not include
contributions from emission lines. Therefore, it is necessary to
correct the emission-line contaminations for the magnitudes when
we compare the predicted colors from stellar population models
with the observed ones (as what we will do in Section 5).

We use the magnitudes computed from the galaxy spectra to correct
emission-line contaminations for the magnitudes. The magnitude
difference estimated from the spectra before and after removing
the emission lines (denoted as $\Delta M_{em}$) represents the
correction inside the fiber. Assuming the contribution of emission
lines to the total magnitude equals to that in the
fiber\footnote{This assumption may not always be correct. Since
the outer part of spirals has bluer colors, which implies a higher
average specific star formation rate ($SFR/M_*$) than that in the
inner part, our method may {\bf underestimate} the emission-line
contamination. On the other hand, due to the contribution of AGN
in the nuclear region of massive galaxies, our method may {\bf
overestimate} the emission-line contamination.}, the sum of
Petrosian magnitude and $\Delta M_{em}$ can be regarded as
emission-free magnitude and the colors estimated from this set of
magnitudes are treated as emission-line-free colors.

Fig. 3 presents distributions of emission-line corrections for
magnitudes and colors in the fiber. It can be seen that, although
the corrections for the magnitudes are small, emission lines do
influence the colors.

\subsection{dust corrections}

Although dust attenuation in face-on late-type galaxy is not
expected to be large, it may have significant effect on the
colors. It would be helpful if the influence of dust can be
estimated and accounted for.

\begin{figure} [t]
\psfig{file=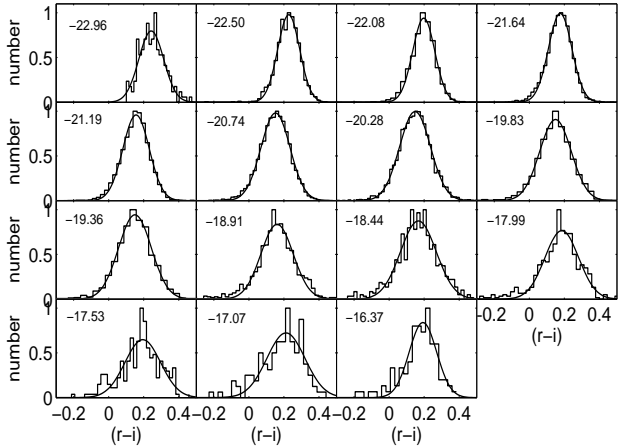,height=6.cm,width=0.5\textwidth}
\caption{Histograms of $r-i$ (after emission-line and dust
corrections) for galaxies in Sample A at different bins. The solid
curves are the best-fit results for Gaussian functions. The median
value of $M_r$ in each bin is shown as the number in each panel.}
\end{figure}

\begin{table*}
\caption{color-magnitude relations} \centering
\begin{tabular}{lrrrrrr|lrrrrrr}
\noalign{\smallskip} \hline \noalign{\smallskip}  &
\multicolumn{2}{c}{$g-r$} & \multicolumn{2}{c}{$r-i$} &
\multicolumn{2}{c}{$r-z$} &  & \multicolumn{2}{c}{$r-J$} &
\multicolumn{2}{c}{$r-K$} &
\multicolumn{2}{c}{$J-K$}\\
$M_r$ & $\bar{C}$ & $\sigma$ & $\bar{C}$ & $\sigma$ & $\bar{C}$ & $\sigma$ & $M_r$& $\bar{C}$ & $\sigma$ & $\bar{C}$ & $\sigma$ & $\bar{C}$ & $\sigma$\\
 (1) & (2) & (3) & (4) & (5) & (6) & (7) & (8) & (9) & (10) & (11) & (12) & (13) & (14) \\
\noalign{\smallskip} \hline \noalign{\smallskip}
-16.37 & 0.42 & 0.12 & 0.19 & 0.08 & 0.31 & 0.16  & -18.58 & 1.52 & 0.24 & 2.46 & 0.29 & 0.91 & 0.18 \\
-17.07 & 0.45 & 0.13 & 0.21 & 0.11 & 0.33 & 0.22  & -19.08 & 1.49 & 0.19 & 2.35 & 0.26 & 0.86 & 0.19 \\
-17.53 & 0.42 & 0.10 & 0.19 & 0.10 & 0.35 & 0.18  & -19.37 & 1.52 & 0.16 & 2.47 & 0.21 & 0.93 & 0.17 \\
-17.99 & 0.41 & 0.09 & 0.19 & 0.10 & 0.31 & 0.20  & -19.66 & 1.52 & 0.17 & 2.44 & 0.21 & 0.92 & 0.13 \\
-18.44 & 0.42 & 0.10 & 0.17 & 0.10 & 0.33 & 0.20  & -19.96 & 1.55 & 0.16 & 2.46 & 0.19 & 0.93 & 0.13 \\
-18.91 & 0.42 & 0.09 & 0.16 & 0.09 & 0.34 & 0.18  & -20.26 & 1.57 & 0.17 & 2.51 & 0.22 & 0.94 & 0.14 \\
-19.36 & 0.43 & 0.08 & 0.15 & 0.09 & 0.33 & 0.17  & -20.57 & 1.57 & 0.16 & 2.53 & 0.20 & 0.94 & 0.13 \\
-19.83 & 0.43 & 0.09 & 0.15 & 0.09 & 0.35 & 0.16  & -20.86 & 1.59 & 0.14 & 2.55 & 0.20 & 0.96 & 0.15 \\
-20.28 & 0.44 & 0.09 & 0.15 & 0.09 & 0.36 & 0.15  & -21.15 & 1.59 & 0.13 & 2.55 & 0.18 & 0.95 & 0.14 \\
-20.74 & 0.43 & 0.09 & 0.15 & 0.08 & 0.36 & 0.14  & -21.45 & 1.61 & 0.14 & 2.59 & 0.18 & 0.97 & 0.14 \\
-21.19 & 0.43 & 0.08 & 0.16 & 0.08 & 0.37 & 0.13  & -21.75 & 1.65 & 0.12 & 2.62 & 0.16 & 0.98 & 0.13 \\
-21.64 & 0.44 & 0.07 & 0.18 & 0.07 & 0.41 & 0.13  & -22.05 & 1.70 & 0.13 & 2.69 & 0.16 & 0.98 & 0.12 \\
-22.08 & 0.47 & 0.08 & 0.20 & 0.06 & 0.45 & 0.13  & -22.34 & 1.76 & 0.13 & 2.76 & 0.16 & 0.99 & 0.14 \\
-22.50 & 0.50 & 0.07 & 0.23 & 0.06 & 0.51 & 0.13  & -22.63 & 1.81 & 0.13 & 2.82 & 0.14 & 1.00 & 0.12 \\
-22.96 & 0.60 & 0.08 & 0.29 & 0.07 & 0.62 & 0.12  & -22.96 & 1.87 & 0.13 & 2.86 & 0.14 & 0.99 & 0.12 \\
\noalign{\smallskip} \hline \noalign{\smallskip}
\end{tabular}
\end{table*}

\begin{figure*}[t]
\psfig{file=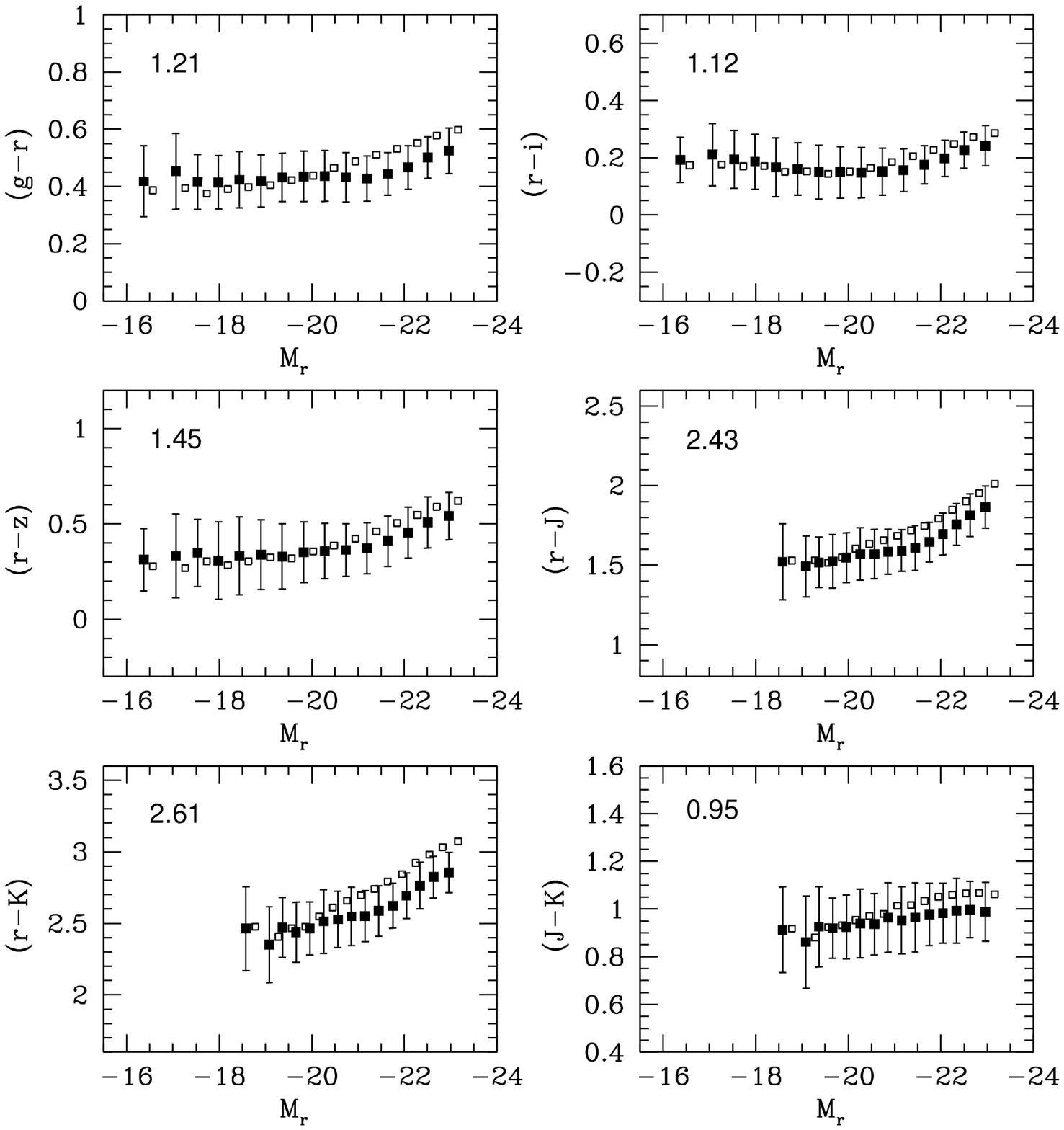,height=12.cm,width=0.9\textwidth}
\caption{Color-magnitude relations, where  the filled square
represents the value for the emission-line-free and dust-free
color (y-axis) as a function of $M_r$ (x-axis) at each bin. The
error bar represents the dispersion $\sigma(M_r)$. Open squares
without error bars are the results after emission-line corrections
but without dust corrections. The x-axis of open square is
slightly shifted for clearness. The number in the upper-left
corner of each panel shows the result of $S=\frac{\Delta
I}{\sigma}$ for each correlation (see text for details).}
\end{figure*}

Kauffmann et al. (2003a) has already obtained the dust attenuation
in the z-band, $A_z$, for SDSS galaxies. They use the $4000 \AA$
break strength and the Balmer absorption line index $H\delta_A$ to
constrain the best star formation history for one galaxy. The dust
extinction is estimated by comparing observed imaging $g-r$ and
$r-i$ colors to model-predicted colors. We adopt $A_z$ from
http://www.mpa-garching.mpg.de/SDSS for our sample galaxies. The
median value of $A_z$ for Sample A and B is 0.21 mag and 0.24 mag,
respectively. By adopting the standard attenuation curve of the
form $\tau_{\lambda} \propto \lambda^{-0.7}$, we can estimate the
dust attenuation in other bands. We refer "colors" as these
emission-line-free and dust-free colors throughout this paper if
there is no special notation.

It should be pointed out that our method to correct the effect of
dust attenuation may not be accurate for individual galaxy.
However, the results for the global trends are robust. Indeed,
Obric et al. (2006) find evidences to confirm that the estimates
of $A_z$ are related to the galaxy dust content.

\section{Results}

\subsection{Color-magnitude relations}

In this section, we use Sample A (only from SDSS) to investigate
optical CMRs and Sample B (a matched sample of SDSS and 2MASS) for
optical-infrared and infrared CMRs. The optical colors ($g-r$,
$r-i$, $r-z$) are estimated from the Petrosian magnitudes. The
optical-infrared colors ($r-J$, $r-K$) are derived using the
aperture-corrected magnitudes in the r-band and the isophotal
fiducial magnitudes from 2MASS. The infrared color $J-K$ are
estimated from the isophotal fiducial magnitudes from 2MASS. All
the magnitudes are corrected for Galactic foreground extinction
and K-correction.

\begin{figure*}[t]
\psfig{file=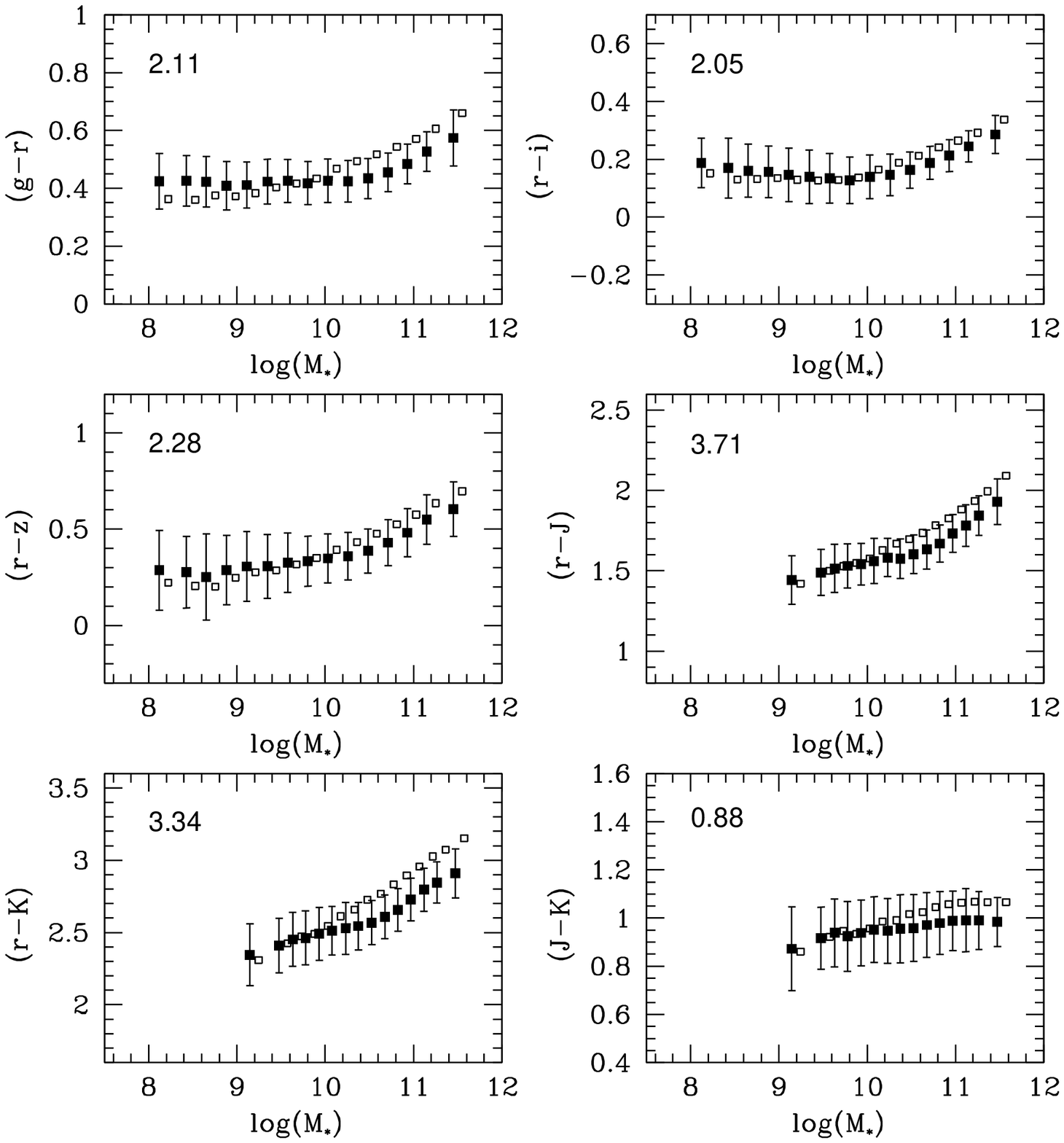,height=12.cm,width=0.9\textwidth}
\caption{Correlations between colors and stellar mass. The caption
is the same as that of Fig. 5.}
\end{figure*}

The galaxies in each sample are divided into 15 bins according to
their absolute magnitudes in the r-band, $M_r$. We choose equal
intervals of $M_r$ in each bin and make sure that each bin
contains more than 100 galaxies. The colors are corrected for dust
extinction by calculating a median value of $A_z$ in each bin
since the estimated $A_z$ for individual galaxy is model-dependent
but the general trend is reliable. It is found that the median
$A_z$ in luminous bin is larger than less luminous bin and the
median $A_z$ is slightly negative for faint galaxies ($M_r>-19$).

Fig. 4 plots histograms of $r-i$ (after emission-line and dust
corrections) for galaxies in Sample A at different bins. The
median value of $M_r$ in each bin is shown as the number in each
panel. It is clear that the distributions of $r-i$ can be well
approximated by Gaussian functions, which are plotted by solid
curves obtained from maximum likelihood estimations. The
distributions of other colors show similar behaviors. Therefore,
we can assume that the color distributions in each bin can be
expressed by a Gaussian function
\begin{equation}
P[C(M_r)]=\frac{1}{2\pi\sigma(M_r)}
exp\{-\frac{[C(M_r)-\bar{C}(M_r)]^2}{2\sigma^2(M_r)}\},
\end{equation}
which is characterized by the mean $\bar{C}(M_r)$ and the
dispersion $\sigma(M_r)$. The best estimations for $\bar{C}(M_r)$
and $\sigma(M_r)$ are listed in Table 1. Further tests show that
main results do not change if a narrower redshift range (such as
$0.01<z<0.1$) is used to construct the sample.

The results for CMRs are plotted in Fig. 5, where the filled
square represents the value of the color (y-axis) as a function of
$M_r$ (x-axis) in each bin, the error bar represents the
dispersion $\sigma(M_r)$. We also show the results after
emission-line corrections but without dust corrections as open
squares without error bars. To quantify the tightness of the
relation, we define a parameter $S=\frac{\Delta I}{\sigma}$, where
$\Delta I$ is the difference between the mean color of the
brightest bin and the faintest bin, $\sigma$ is the average value
for the color dispersions in 15 bins. Thus, higher $S$ corresponds
to tighter relation. The resulted $S$ for the CMR is shown in the
upper-left corner of each panel in Fig. 5.

It can be seen that dust attenuation contributes significantly to
the observed CMRs. This is expected since luminous galaxy is more
attenuated and reddened than less luminous object (Tully et al.
1998; Masters et al. 2003). After correcting the dust attenuation,
the optical colors ($g-r$, $r-i$, $r-z$) for faint galaxies (i.e.
$M_r> -21$) have very weak correlation with the luminosity, while
the optical colors for bright galaxies (i.e. $M_r < -21$) are
redder for more luminous galaxies. This is consistent with the
results of Baldry et al. (2004), who also find a shallow CMR slope
for the low-luminosity blue-distribution galaxies.

It is also shown that, after correcting dust attenuation, the
optical-infrared ($r-J$, $r-K$) CMRs are still very tight. For
infrared color ($J-K$), a weak CMR is presented, in the sense that
the color range $\Delta I$ is comparable with the mean dispersion
$\sigma$. Unfortunately, due to the fact that 2MASS is much more
shallow than SDSS, there is no sufficient number of less luminous
galaxies in Sample B to give the optical-infrared and infrared
CMRs behavior at the fainter end.

\subsection{Correlations between colors and stellar mass}

Stellar mass is another fundamental physical parameter for
galaxies. Now, we turn to investigate the color-mass relations for
late-type galaxies.

Again, we divide the galaxies into 15 bins according to their
stellar mass $M_*$ (which is in units of $M_{\odot}$ throughout
this paper) and fit the distribution of the color in each bin by a
Gaussian function. The method to correct dust attenuation is the
same as that in section 4.1. The best-fit mean of color and its
dispersion in each bin are listed in Table 2 and plotted in Fig.
6. The most remarkable feature of Fig. 6 is the tightness of
correlation between optical-infrared color ($r-J$, $r-K$) and
stellar mass, in the sense that $\Delta I$ is nearly (if not more
than) more than 3 times the mean color dispersion. We will discuss
the implication of these results in Section 5.

It could be interesting to compare the CMRs with color-mass
relations. The colors show similar but stronger correlations with
stellar mass than luminosity. The colors in the reddest bins are
redder and the colors in the bluest bins are bluer when mass is
used instead of absolute magnitude. This is true for all
wavelength bands. However, it's not completely clear how much of
the steeper slopes and tighter scatter in color-mass relations
relative to CMRs is driven by the color dependence of stellar
$M/L$ values. Tremonti et al. (2004) also find that, when
luminosity is used instead of mass, both dust and $M/L$ variations
act to smear out the turnover seen in the mass-metallicity
relation at high metallicity.

\begin{table*}
\caption{Correlations between colors and stellar mass.} \centering
\begin{tabular}{lrrrrrr|lrrrrrr}
\noalign{\smallskip} \hline \noalign{\smallskip}  &
\multicolumn{2}{c}{$g-r$} & \multicolumn{2}{c}{$r-i$} &
\multicolumn{2}{c}{$r-z$} &  & \multicolumn{2}{c}{$r-J$} &
\multicolumn{2}{c}{$r-K$} &
\multicolumn{2}{c}{$J-K$}\\
$\log(M_*)$ & $\bar{C}$ & $\sigma$ & $\bar{C}$ & $\sigma$ & $\bar{C}$ & $\sigma$ & $\log(M_*)$ & $\bar{C}$ & $\sigma$ & $\bar{C}$ & $\sigma$ & $\bar{C}$ & $\sigma$\\
 (1) & (2) & (3) & (4) & (5) & (6) & (7) & (8) & (9) & (10) & (11) & (12) & (13) & (14) \\
\noalign{\smallskip} \hline \noalign{\smallskip}
8.12 & 0.42 & 0.10 & 0.19 & 0.09 & 0.29 & 0.21 & 9.15 & 1.44 & 0.15 & 2.35 & 0.21 & 0.87 & 0.17 \\
8.43 & 0.43 & 0.09 & 0.17 & 0.10 & 0.28 & 0.19 & 9.48 & 1.49 & 0.14 & 2.41 & 0.19 & 0.92 & 0.13 \\
8.65 & 0.42 & 0.09 & 0.16 & 0.09 & 0.25 & 0.22 & 9.63 & 1.51 & 0.15 & 2.45 & 0.19 & 0.94 & 0.14 \\
8.88 & 0.41 & 0.08 & 0.16 & 0.09 & 0.29 & 0.18 & 9.78 & 1.53 & 0.14 & 2.46 & 0.19 & 0.92 & 0.14 \\
9.11 & 0.41 & 0.08 & 0.15 & 0.09 & 0.31 & 0.18 & 9.93 & 1.54 & 0.13 & 2.49 & 0.18 & 0.94 & 0.14 \\
9.35 & 0.42 & 0.08 & 0.14 & 0.09 & 0.31 & 0.17 & 10.08 & 1.56 & 0.14 & 2.51 & 0.17 & 0.95 & 0.14 \\
9.57 & 0.43 & 0.07 & 0.13 & 0.09 & 0.33 & 0.15 & 10.23 & 1.58 & 0.12 & 2.53 & 0.18 & 0.95 & 0.14 \\
9.80 & 0.42 & 0.07 & 0.13 & 0.08 & 0.33 & 0.13 & 10.38 & 1.58 & 0.12 & 2.54 & 0.16 & 0.95 & 0.14 \\
10.03 & 0.43 & 0.08 & 0.14 & 0.08 & 0.35 & 0.13 & 10.53 & 1.60 & 0.12 & 2.57 & 0.15 & 0.96 & 0.14 \\
10.26 & 0.42 & 0.07 & 0.15 & 0.07 & 0.36 & 0.12 & 10.67 & 1.63 & 0.12 & 2.61 & 0.15 & 0.97 & 0.13 \\
10.48 & 0.43 & 0.07 & 0.16 & 0.06 & 0.39 & 0.11 & 10.82 & 1.67 & 0.12 & 2.66 & 0.15 & 0.98 & 0.13 \\
10.71 & 0.46 & 0.07 & 0.19 & 0.06 & 0.43 & 0.12 & 10.97 & 1.73 & 0.12 & 2.73 & 0.15 & 0.99 & 0.12 \\
10.93 & 0.48 & 0.07 & 0.21 & 0.06 & 0.48 & 0.13 & 11.12 & 1.78 & 0.13 & 2.80 & 0.15 & 0.99 & 0.13 \\
11.15 & 0.53 & 0.07 & 0.25 & 0.05 & 0.55 & 0.13 & 11.27 & 1.84 & 0.12 & 2.85 & 0.14 & 0.99 & 0.12 \\
11.45 & 0.57 & 0.10 & 0.29 & 0.07 & 0.60 & 0.14 & 11.47 & 1.93 & 0.14 & 2.91 & 0.17 & 0.98 & 0.10 \\
 \noalign{\smallskip} \hline
\noalign{\smallskip}
\end{tabular}
\end{table*}

\begin{figure*} [t]
\psfig{file=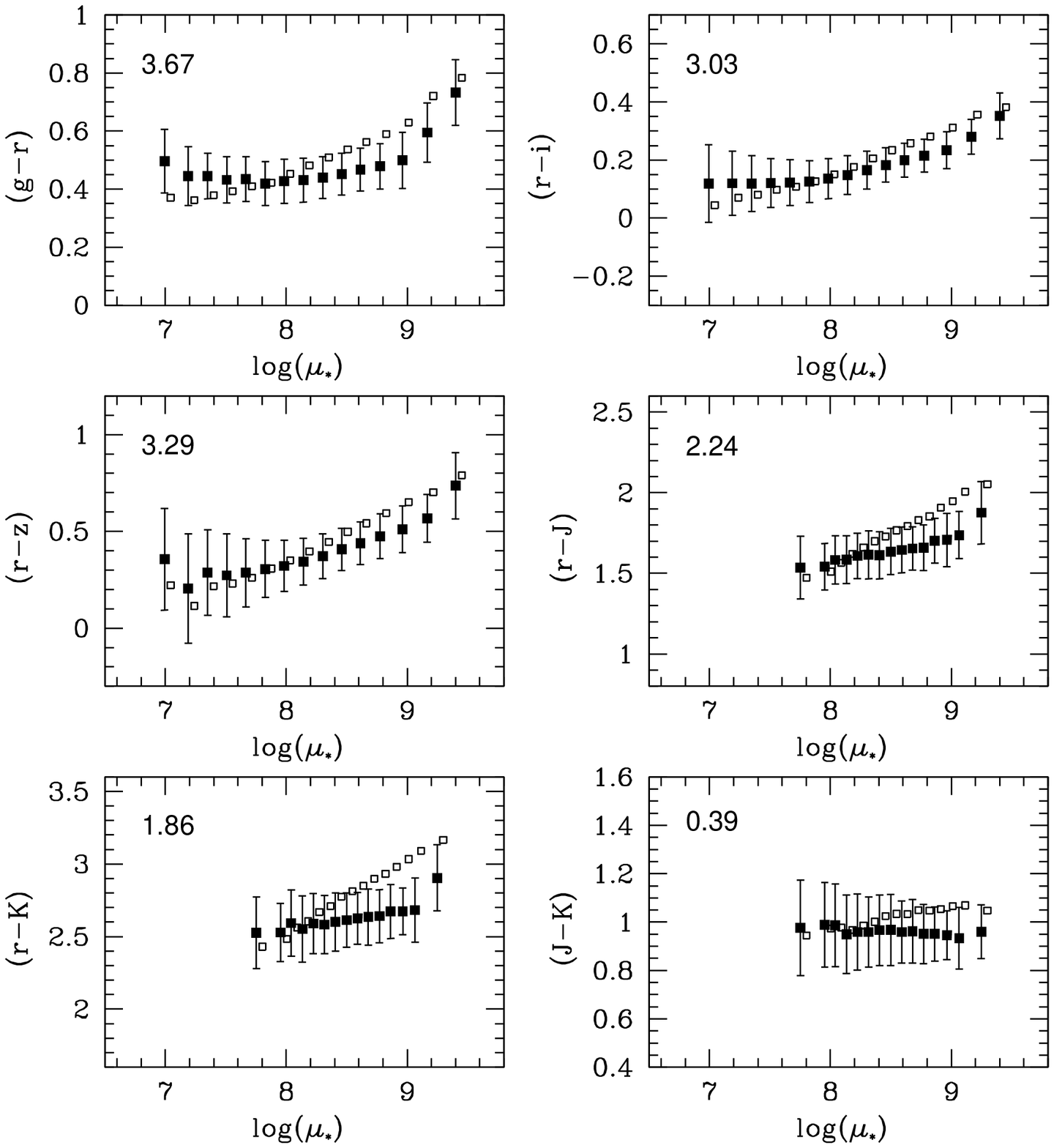,height=12.cm,width=0.9\textwidth}
\caption{Correlations between colors and stellar mass surface
density. The caption is the same as that of Fig. 5. }
\end{figure*}

\subsection{Correlations between colors and mass surface density}

Stellar mass surface density, which describes how luminous mass is
distributed along the disk, is another fundamental parameter for
spiral galaxies. The half-light radius in the $z$-band and stellar
mass yield the effective stellar mass surface density $\mu_* =
M_*/2 \pi R_{50,z}^2$ (Kauffmann et al. 2005; Chang et al. 2006),
which is in units of $\rm{M_{\odot}kpc^{-2}}$. The results for
color-$\mu_*$ relations are given in Table 3 and Fig. 7.

\begin{table*}
\caption{Correlations between colors and stellar surface mass
density $\log(\mu_*)$} \centering
\begin{tabular}{lrrrrrr|lrrrrrr}
\noalign{\smallskip} \hline \noalign{\smallskip}  &
\multicolumn{2}{c}{$g-r$} & \multicolumn{2}{c}{$r-i$} &
\multicolumn{2}{c}{$r-z$} &  & \multicolumn{2}{c}{$r-J$} &
\multicolumn{2}{c}{$r-K$} &
\multicolumn{2}{c}{$J-K$}\\
$\log(\mu_*)$ & $\bar{C}$ & $\sigma$ & $\bar{C}$ & $\sigma$ & $\bar{C}$ & $\sigma$ & $\log(\mu_*)$& $\bar{C}$ & $\sigma$ & $\bar{C}$ & $\sigma$ & $\bar{C}$ & $\sigma$\\
 (1) & (2) & (3) & (4) & (5) & (6) & (7) & (8) & (9) & (10) & (11) & (12) & (13) & (14) \\
\noalign{\smallskip} \hline \noalign{\smallskip}
6.99 & 0.50 & 0.11 & 0.12 & 0.13 & 0.36 & 0.26 & 7.75 & 1.53 & 0.19 & 2.53 & 0.25 & 0.98 & 0.20  \\
7.19 & 0.45 & 0.10 & 0.12 & 0.11 & 0.20 & 0.28 & 7.95 & 1.54 & 0.14 & 2.53 & 0.20 & 0.99 & 0.18 \\
7.35 & 0.44 & 0.08 & 0.12 & 0.10 & 0.29 & 0.22 & 8.04 & 1.58 & 0.15 & 2.59 & 0.23 & 0.99 & 0.17 \\
7.51 & 0.43 & 0.08 & 0.12 & 0.08 & 0.27 & 0.21 & 8.13 & 1.58 & 0.15 & 2.55 & 0.23 & 0.95 & 0.16 \\
7.67 & 0.43 & 0.08 & 0.12 & 0.08 & 0.29 & 0.18 & 8.22 & 1.61 & 0.14 & 2.59 & 0.21 & 0.96 & 0.16 \\
7.83 & 0.42 & 0.08 & 0.13 & 0.07 & 0.31 & 0.15 & 8.32 & 1.61 & 0.15 & 2.58 & 0.20 & 0.96 & 0.15 \\
7.98 & 0.43 & 0.08 & 0.14 & 0.07 & 0.32 & 0.13 & 8.41 & 1.61 & 0.15 & 2.60 & 0.20 & 0.97 & 0.15 \\
8.14 & 0.43 & 0.08 & 0.15 & 0.07 & 0.34 & 0.12 & 8.50 & 1.64 & 0.14 & 2.61 & 0.19 & 0.97 & 0.15 \\
8.30 & 0.44 & 0.07 & 0.16 & 0.06 & 0.37 & 0.11 & 8.59 & 1.64 & 0.14 & 2.63 & 0.18 & 0.96 & 0.13 \\
8.46 & 0.45 & 0.07 & 0.18 & 0.06 & 0.40 & 0.11 & 8.68 & 1.65 & 0.13 & 2.64 & 0.19 & 0.96 & 0.13 \\
8.61 & 0.47 & 0.07 & 0.20 & 0.06 & 0.44 & 0.11 & 8.77 & 1.66 & 0.14 & 2.64 & 0.18 & 0.95 & 0.12 \\
8.78 & 0.48 & 0.08 & 0.21 & 0.06 & 0.47 & 0.12 & 8.86 & 1.70 & 0.14 & 2.67 & 0.19 & 0.95 & 0.11 \\
8.96 & 0.50 & 0.10 & 0.23 & 0.06 & 0.51 & 0.12 & 8.96 & 1.71 & 0.16 & 2.67 & 0.16 & 0.95 & 0.10 \\
9.17 & 0.59 & 0.10 & 0.28 & 0.06 & 0.57 & 0.12 & 9.07 & 1.74 & 0.15 & 2.68 & 0.22 & 0.93 & 0.13 \\
9.40 & 0.73 & 0.11 & 0.35 & 0.08 & 0.74 & 0.17 & 9.25 & 1.88 & 0.19 & 2.90 & 0.23 & 0.96 & 0.11 \\
\noalign{\smallskip} \hline \noalign{\smallskip}
\end{tabular}
\end{table*}

It can be seen that, both optical and optical-infrared colors show
correlations with stellar mass surface density, i.e. galaxies with
higher $\mu_*$ have redder colors. This point is anyhow expected
since high mass surface density corresponds to higher stellar mass
for late-type galaxies (Shen et al. 2003). More interestingly,
when $\mu_*$ is used in place of $M_*$, the resulted S increases
for optical color, while S decreases for optical-infrared and
infrared colors. In other words, for late-type galaxies, the
optical colors correlate more strongly with mass surface density
than with stellar mass, while optical-infrared and infrared colors
show stronger correlations with stellar mass.

This is obviously different from what is obtained for early-type
galaxies. Chang et al. (2006) have shown that, for elliptical
galaxies, all galaxy colors correlate more strongly with stellar
mass than with mass surface density, which suggests that the star
formation history of nearby elliptical galaxies is primarily
determined by the mass of the system. Since the age-metallicity
degeneracy makes it difficult to directly deduce the property of
stellar population from colors, it is instructive to break the
age-metallicity degeneracy firstly, such as adopting the SPS
models, and then discuss the implications of various correlations
on the star formation histories (SFHs) for late-type galaxies.

\section{Stellar populations of late-type galaxies}

Motivated by the work of Bell \& de Jong (2000), we use a
combination of optical and near-infrared broad-band colors for
galaxies in our sample B to probe the stellar populations of late
type galaxies. In other words, we generate a grid of model
spectral energy distributions  using SPS models and estimate the
model parameters by matching all the available photometry of the
sample galaxies to the colors predicted by the SPS models. By
doing this, we are able to investigate the trends in the mean
stellar age and metallicity as a function of stellar mass and mass
surface density.

We adopt the SPS models of Bruzual \& Charlot (2003). The
universal initial mass function is taken from Chabrier (2003). The
lower and upper mass cut-offs are set as $m_L$=0.1M$_{\odot}$ and
$m_U$=100M$_{\odot}$, respectively. We use simplified SFHs and
given metallicities to construct a grid of model colors. The star
formation rate (SFR) is parameterized by:
\begin{eqnarray}
\Psi(t) & = & \left \{ \begin{array}{ll} 0 & {\rm if\ } t < t_{form} \\
B t e^{-t/\tau_{SF}} & {\rm if\ } t_{form}\le t\le t_g,
\end{array}
\right.
\end{eqnarray}
where $B$ is a normalization constant which determines the total
mass of the stellar population, $\tau_{SF}$ is the star formation
timescale, $t_g$ is the age of galaxy and is fixed as $t_g=12Gyr$,
$t_{form}$ characterizes the time delay for star formation. Under
the above assumptions, the mean age of the stellar population
$<age>$ is given by:
\begin{equation}
<age>=\frac{\int_{0}^{t_g} (t_g-t) \psi(t)\, {\rm d}t
}{\int_{0}^{t_g} \psi(t)\, {\rm d}t}.
\end{equation}

Therefore, the mean age of stellar population is determined by the
combination of $\tau_{SF}$ and $t_{form}$. If $t_{form}=0$, the
youngest mean age given by none-negative $\tau_{SF}$ is around
4Gyr, which corresponds to $\tau\rightarrow \infty$ and the SFR
increases with time. In this case, the stellar population is still
too old to fit the colors of some observed blue galaxies. As an
alternative, we assume a none-zero $t_{form}$, which insures that
our models also include the youngest stellar populations. We
assume $t_{form}=0.2 \tau_{SF}$ during model calculations. This
assumption is acceptable for two reasons. First, we focus only on
the mean stellar age rather than details of SFH for each
individual galaxy. Fixing the relation between $\tau_{SF}$ and
$t_{form}$ will reduce the number of free parameters in the model.
Second, SPS model has the advantage of giving robust relative ages
and metallicities estimations (see discussions in Bell \& de Jong
2000). We will emphasize the {\it relative trends} rather than the
absolute values of ages and metallicities. Indeed, test
calculations show that the main conclusions are not sensitive to
either the adopted form of SFH or the assumption $t_{form}=0.2
\tau_{SF}$.

Thus, there are two free parameters in the model: metallicity and
$\tau_{SF}$. $\tau_{SF}$ progresses from 0 to $60 Gyr$
($\tau_{SF}=60$Gyr corresponds to the upper limit for
$t_{form}=12Gyr$) smoothly. This results in the mean age of
stellar population going from old to young. Firstly, we generate a
model grid of single-burst stellar population (SSPs) for different
metallicity by interpolating along the model metallicities. The
range of the model metallicities is from $0.005Z_{\odot}$ to
$2.5Z_{\odot}$. Then, we generate a model grid for a number of
$\tau_{SF}$ values by integrating the SSPs with a given
metallicity. These models provide a set of model colors
$C_{model,i} (<age>,Z)$ (where $i$ denotes different color) for a
range of SFHs and metallicity.

We match all the available colors of the galaxies to the colors
predicted by the SPS models and estimate the model parameters
using $\chi^2$-fitting. For each $\tau_{SF}$ and metallicity $Z$,
the corresponding $\chi^2$ is given by:
\begin{equation}
\chi^2=\sum_{i=1}^{n}\frac{(C_{model,i}-C_{obs,i})^2}{\sigma_i^2}.
\end{equation}
where $\sigma_i$ is the observed error for color $i$. The best
model match $(<age>,Z)$ is the one with the minimum $\chi^2$
value.

We use the observed $g-r$, $r-i$, $r-z$, $r-J$ and $r-K$ colors in
$\chi^2$-fitting. These colors are corrected by emission-line
contamination and dust attenuation. Since the estimated $A_z$ is
model-dependent and may not be accurate for individual galaxy, we
choose to correct an average trend of dust attenuation by fitting
a linear relation between $A_z$ and $M_*$, i.e. to estimate $A_z$
from stellar mass. In general, massive galaxies have larger dust
attenuation. The results for $\chi^2$-fitting show that there are
about 5 percent of galaxies with the minimum $\chi^2$ value
located in either boundary of  the model parameters $\tau_{SF}$
and $Z$. We exclude these galaxies from further analysis below.

\begin{figure*} [t]
\psfig{file=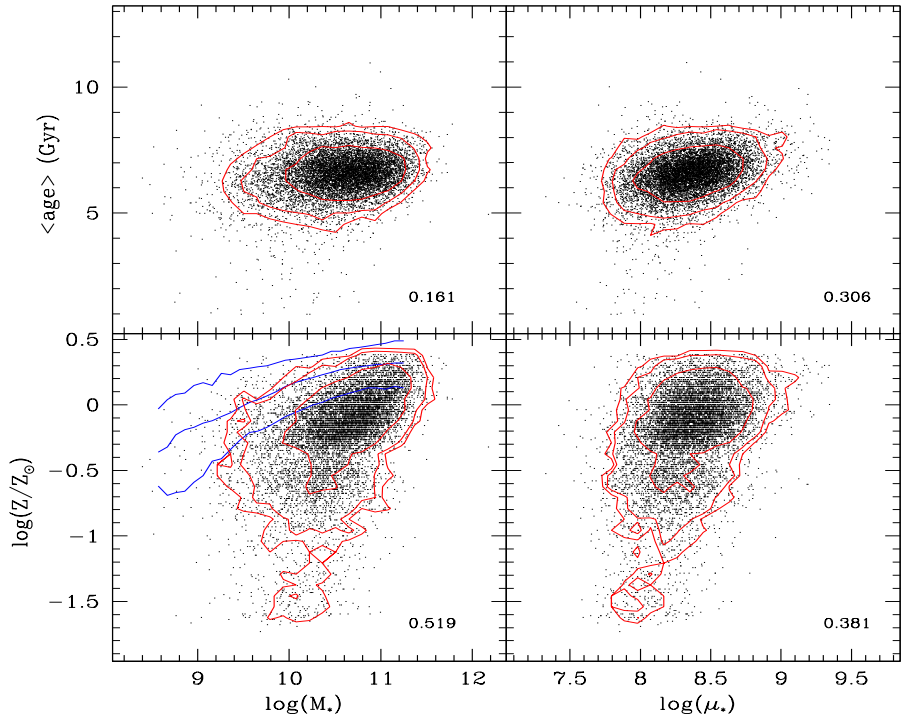,height=11.cm,width=0.8\textwidth}
\caption{Model predictions of mean age ($<age>$) and metallicity
($Z$) against stellar mass and mass surface density for galaxies
in Sample B. The set of contours indicates the positions within
which how much fraction of total number of galaxies is contained.
The levels of contours correspond to $68\%$, $90\%$ and $95\%$,
respectively. The number in the right corner of each panel shows
the Spearman rank-order correlation coefficient. Three blue lines
in bottom-left panel plot the observed rang for the relation
between stellar mass and gas-phase metallicity, which is taken
from Table 3 in Tremonti et al.(2004). The solar $12+log(O/H)$ is
taken as 8.8 (Grevesse \& Sauval 1998).}
\end{figure*}

Fig. 8 plots the model predictions of $(<age>,Z)$ against stellar
mass and mass surface density for galaxies in Sample B. To show
the distribution more clearly, we plot sets of contours to
indicate the positions within which different fraction of total
number of galaxies is contained. We also adopt the Spearman
rank-order test to quantify the significance of the relation and
show the resulted correlation coefficient in the right corner of
each panel. Fig. 8 shows that massive galaxies have older and
higher metallicity stellar populations than less massive galaxies.
In other words, not only the stellar metallicity but also the
stellar mean age that contribute to the observed CMRs for
late-type galaxies. To contrast, the CMRs of early-type galaxies
are mainly ascribed to the metallicity effect: more luminous
galaxies have redder colors because they have higher metallicities
(Faber 1977; Dressler 1984; Arimoto \& Yoshii 1987; Kodama \&
Arimoto 1997; Kauffmann \& Charlot 1998; De Lucia et al. 2004;
Kang et al. 2005).

The general trend for mass-metallicity relation found in this
paper is similar to some earlier works (such as Peltier \& de
Grijs 1998; Bell \& de Jong 2000). Moreover, gas-phase oxygen
abundance in star-forming galaxy has already been found to be
related to the luminosity or/and stellar mass of the system
(Kobulnicky \& Zaritsky 1999; Kobulnivky \& Kewley 2004; Hammer et
al. 2005; Liang et al. 2004; 2006 etc.). For comparison, we show
the observed mass-metallicity relation presented by Tremonti et
al.(2004) with three blue lines. To transfer $12+log(O/H)$ to
$log(Z/Z_{\odot})$, we assume the metallicity $Z$ can be
represented by oxygen abundance and take the solar value as 8.8
(Grevesse \& Sauval 1998). Since the metallicity derived from SPS
models is the mean value of the stellar population, it is
reasonable that the model predicted mettallicty is somewhat lower
than that of Tremonti et al. (2004). As we can see, the general
trend of our results deduced from the CMRs are consistent with
their independent measurements from emission lines.

More interestingly, it can be seen (even by eye) that the
metallicity correlates more strongly with $M_*$ than $\mu_*$. We
also find that the residuals from the age-$\mu_*$ relation do not
correlate with $M_*$, while the residuals from the age-$M_*$
relation increase with $\mu_*$. All these suggest that the stellar
mass of late-type galaxy is the primary parameter to determine the
mean metallicity of the system. This is probably, similar as
early-type galaxy, because the stellar mass strongly correlates to
the potential well of the galaxy and then determines the
efficiency for ejecting the metals of the system. Tremonti et al.
(2004) also interpret their results as strong evidence of both the
ubiquity of galactic winds and their effectiveness in removing
metals from galaxy potential wells.

On the other hand, the resulted mean stellar age $<age>$ is more
sensitive to $\mu_*$ than $M_*$. It is also found that the
residuals from the metallicity-$\mu_*$ relation strongly correlate
with $M_*$, while the residuals from the metallicity-$M_*$
relation have no correlation with $\mu_*$. In other words, our
results suggest that mass surface density plays an more important
role in regulating the star formation history of late-type galaxy
than stellar mass. Indeed, some early studies already pointed out
that the star formation history for late-type galaxies correlates
more strongly with mass surface density than stellar mass (Bell \&
de Jong 2000; Kauffmann et al. 2003b; 2006). Kennicutt (1998) also
shows that it is local (e.g. surface density) rather than global
factors (e.g. stellar mass) that regulate the rate at which spiral
galaxies form their stars at the present day.

\section{Summary}

We have collected two samples of face-on late-type galaxies to
investigate the correlations between colors and absolute
magnitude, stellar mass and stellar mass surface density. Sample A
is selected from DR4 of SDSS and Sample B is a position matched
sample between SDSS (DR4) and 2MASS. In order to correct the
aperture mismatch between SDSS and 2MASS, we use the radial
profile of SDSS galaxies and correct the SDSS magnitudes to the
isophotal circular radius where the 2MASS magnitudes are measured.
The emission-line contamination is corrected for the magnitudes by
comparing the magnitudes computed from the spectra before and
after removing the emission lines. We also correct dust
attenuation by adopting $A_z$ estimation plus an standard kind of
attenuation curve. Moreover, we match the observed colors of our
sample to the colors predicted by stellar population synthesis
model and investigate their implications of the stellar
populations. We find that:

\begin{enumerate}

\item The optical colors for faint galaxies (i.e. $M_r
> -21$) have very weak correlation with the luminosity, while the
optical colors for bright galaxies (i.e. $M_r < -21$) are redder
for more luminous galaxies. The optical-infrared colors show
tighter CMRs than optical ones.

\item All (optical, optical-infrared and infrared) colors show
similar but stronger correlations with stellar mass than with
absolute magnitudes.

\item The optical colors correlate more strongly with mass surface
density than with stellar mass, while optical-infrared and
infrared colors show stronger correlations with stellar mass.

\item By comparing the observed colors with SPS model colors, we
find that massive galaxies have older and higher metallicity
stellar populations than less massive galaxies. This suggests that
the CMRs of late-type galaxies are the combined results of stellar
mean age and metallicity.

\item It is also shown that the mean stellar metallicity is more
sensitive to $M_*$ than to $\mu_*$, while the mean stellar age is
more sensitive to $\mu_*$ than $M_*$. This suggests that the
stellar metallicity is mainly determined by the mass of the
system, while the star formation history of late-type galaxy is
mainly regulated by mass surface density.
\end{enumerate}

Our results are based on the homogeneous and large galaxies
samples from SDSS and 2MASS to derive the CMRs for late-type
galaxies in both optical and optical-infrared bands. Since the
combination of optical and near-infrared broad-band colors
contains more information on the star formation process, our
observed results have provided strong constraints on the model
descriptions for gas accretion and supernova feedback processes in
late-type galaxies. Further comparisons between model predictions
and observed data are needed to investigate the star formation
histories for late-type galaxies.

\section*{Acknowledgements}

The authors thank the anonymous referee for his/her helpful
suggestions to greatly improve this paper. We would like to thank
Houjun Mo, Shude Mao, Guinevere Kauffmann, Eric Peng and
St\'ephane Charlot for useful discussions. We thank Jarle
Brinchmann for providing the data of spectral magnitudes. This
project is partly supported by NSFC 10403008, 10573028, 10333020,
10173017, NKBRSF 1999075404, Shanghai Municipal Science and
Technology Commission No. 04dz05905.

Funding for the creation and distribution of the SDSS Archive has
been provided by the Alfred P. Sloan Foundation, the Participating
Institutions, the National Aeronautics and Space Administration,
the National Science Foundation, the U.S. Department of Energy,
the Japanese Monbukagakusho, and the Max Planck Society. The SDSS
Web site is http://www.sdss.org/.

The SDSS is managed by the Astrophysical Research Consortium (ARC)
for the Participating Institutions. The Participating Institutions
are The University of Chicago, Fermilab, the Institute for
Advanced Study, the Japan Participation Group, The Johns Hopkins
University, Los Alamos National Laboratory, the
Max-Planck-Institute for Astronomy (MPIA), the
Max-Planck-Institute for Astrophysics (MPA), New Mexico State
University, University of Pittsburgh, Princeton University, the
United States Naval Observatory, and the University of Washington.

\end{document}